# Ferroelectric Switching in Hybrid Molecular Beam Epitaxy-Grown BaTiO$_3$ Films

Anusha Kamath Manjeshwar[1,*], Zhifei Yang[1,2], Chin-Hsiang Liao[3], Jiaxuan Wen[3], Steven J. Koester[3], Richard D. James[4], and Bharat Jalan[1,*]

[1]Department of Chemical Engineering and Materials Science, University of Minnesota, Minneapolis, Minnesota, United States – 55455

[2]School of Physics and Astronomy, University of Minnesota, Minneapolis, Minnesota, United States – 55455

[3]Department of Electrical and Computer Engineering, University of Minnesota, Minneapolis, Minnesota, United States – 55455

[4]Department of Aerospace Engineering and Mechanics, University of Minnesota, Minneapolis, Minnesota, United States – 55455

* Corresponding authors: Anusha Kamath Manjeshwar, kamat086@umn.edu; Bharat Jalan, bjalan@umn.edu

**Abstract**

Molecular beam epitaxy (MBE) is a promising synthesis technique for both heterostructure growth and epitaxial integration of ferroelectric BaTiO$_3$. However, a direct measurement of the remnant polarization ($P_r$) has not been previously reported in MBE-grown BaTiO$_3$ films. We report the *in-situ* growth of an all-epitaxial SrRuO$_3$/BaTiO$_3$/SrRuO$_3$ heterostructure on Nb-doped SrTiO$_3$ (001) substrates by hybrid MBE using metal-organic precursors. This capacitor structure consisting of 16 nm SrRuO$_3$/40 nm BaTiO$_3$/16 nm SrRuO$_3$ shows hysteretic polarization-electric field (*P-E*) curves with $P_r \sim 15$ µC cm$^{-2}$ at frequencies ranging from 500 Hz to 20 kHz, after isolating the intrinsic ferroelectric response from non-ferroelectric contributions using the Positive-Up-Negative-Down (PUND) method. We hypothesize that the asymmetry in switching behavior and current leakage has origins in structural defects.

## 1. Introduction

Barium titanate (BaTiO₃, BTO) is a leading lead-free high-κ dielectric and ferroelectric candidate[1] for applications in non-volatile memories, capacitors, sensors, and electro-optic devices. Bulk BTO has a tetragonal crystal structure at room temperature with lattice constants of $a = b = 3.992$ Å and $c = 4.036$ Å and a Curie temperature, $T_c \sim 120$ °C[2]. BTO also exhibits one of the highest Pockels coefficients in bulk single crystals[3], thin films[4], and when epitaxially integrated on Si[5], making it one of the best suited materials for electro-optical modulators[6], low-power optical switches, BTO-Si waveguides with low propagation losses[7], and photonic racetrack resonators for high-speed optical communication and neuromorphic computing[3,8]. In addition to these electro-optical applications, BTO is promising for harvesting low-grade waste heat[9–12] due to its relatively low Curie temperature and sharp, fatigue-free[13] ferroelectric-to-paraelectric phase transformation with a discontinuous change in polarization.

Strategies inspired by the high-performance metrics of PZT-based materials[14,15] — such as increased tetragonality (ratio of the out-of-plane ($c$ or $a_\perp$) to the in-plane lattice parameters ($a$ or $a_\parallel$)) and engineered phase boundaries (to allow easier polarization rotation among transforming phases) — have been actively pursued to tailor the ferroelectric properties of BTO-based films. Two prominent approaches include the application of compressive strain[16–19] and alloying with Sn[20–23], the latter of which introduces a point of four-phase coexistence in the temperature-composition phase diagram at approximately 11% Sn in Sn-alloyed BTO ceramics[20]. Choi *et al.*[18] demonstrated that BTO films grown on compressively straining substrates such as GdScO₃ (110) and DyScO₃ (110) exhibit an up to 270% enhancement in remnant polarization ($P_r$) relative to bulk single crystals[24]. However, this gain was accompanied by a 24-fold increase in coercive field ($E_c$), indicating a significant trade-off. Complementing this, Jiang *et al.*[25] showed that minimizing depolarization fields in BTO capacitors using SrRuO$_3$ (SRO) electrodes—through careful optimization of growth and device processing conditions—could move past this trade-off, enabling low-voltage polarization switching while maintaining relatively high $P_r$ values and low $E_c$. These examples underscore the importance of epitaxial growth methods and device fabrication protocols in determining ferroelectric switching behavior and overall device performance.

Molecular beam epitaxy (MBE) is a thin-film synthesis technique uniquely suited for thin film heterostructures due to the low energy of deposition and the ability to produce atomically sharp interfaces. McKee *et al*.[26] pioneered the epitaxial integration of functional perovskite oxides, including BTO, on silicon substrates through the direct growth of high-quality SrTiO$_3$ (STO) buffer layers with minimal interfacial amorphization using MBE. This breakthrough was enabled by the ultra-high vacuum growth environment and low oxygen background pressures of MBE, which are difficult to achieve with other physical vapor deposition (PVD) methods. MBE has also facilitated the systematic control of *c*-domain orientation to favor out-of-plane polarization, by tuning BTO thickness[27,28], growth temperatures[29], oxygen background pressures during growth[29], post-growth annealing conditions[30], and the use of superlattices[31] and misfit-relaxed buffer layers[19,32]. While the MBE growth of *c*-domain BTO thin films has been extensively studied, the investigation of polarization switching has been limited to indirect measurements such as piezoresponse force microscopy (PFM)[19,29,33–35] and hysteresis in the electro-optic response[5,30,36]. Unlike pulsed laser deposition (PLD) and sputtering, a direct observation of hysteresis in polarization with applied electric field and frequency has been largely missing in MBE-grown BTO films (**Figure 1a**)[18,25,37–70]. To our knowledge, only one prior study has reported hysteretic *P–E* curves in MBE-grown films[37], but measurement of remnant polarization or coercivity were not presented. For a ~ 12 nm-thick BTO on Nb:STO (001), they achieved a room-temperature saturation polarization of ~ 3 µC cm$^{-2}$ at 60 Hz—approximately one-tenth of that reported for similarly thick BTO films grown on SRO-buffered STO (001) substrates by sputtering[38]. This observation points to two synthesis parameters which may favor higher polarization and remanence in sputtered BTO over MBE-grown BTO: (1) the orders-of-magnitude higher oxygen background pressures that lower the density of oxygen vacancies which can prematurely suppress ferroelectricity, and (2) the bottom electrode with a higher Schottky barrier. In this letter, we present the hybrid MBE growth of high-quality, all-epitaxial SRO/BTO/SRO heterostructures on Nb:STO (001) substrates. We demonstrate ferroelectric switching behavior with hysteretic *P-E* curves and a remnant polarization of ~ 15 µC cm$^{-2}$ in a 40-nm thick BTO layer, a first for all-epitaxial metal-ferroelectric BTO-metal heterostructures grown by MBE.

## 2. Results

### 2.1. Rationale for choice of epitaxial electrodes for BaTiO$_3$

Reports of metal-ferroelectric BTO-metal heterostructures with hysteretic *P-E* curves using PVD techniques[18,25,37–70] appear to prefer some electrode candidates over others (**Figure 1b**). We highlight three requirements for electrode candidates in metal–ferroelectric BTO–metal heterostructures to rationalize these choices: (1) high conductivity to minimize depolarization fields, (2) formation of a Schottky barrier with BTO, and (3) structural compatibility and low intermixing with BTO. The Schottky-Mott conditions for metal/*n*-type-semiconductor junctions in equilibrium, as a first approximation, suggest that Schottky junctions require a higher metal work function than the semiconductor electron affinity. Hence, both electrodes must have high work functions so that both metal-BTO junctions are Schottky junctions. **Figure 1c** compiles the work functions of electrodes demonstrated for metal-ferroelectric BTO-metal heterostructures synthesized by different PVD techniques[18,25,37–69,71]. Notably, candidates predominantly chosen as top electrodes exhibit higher work functions on average than the bottom electrode candidates[72–81]. Bottom electrode candidates must meet three additional criteria: (1) they enable *in-situ* epitaxial growth of *c*-domain BTO through compressive misfit strain (**Figure 1d**); (2) their $(001)_{pc}$ (pseudocubic) surfaces promote wetting to avoid 3D island growth, an issue predicted for Pt (001)/BTO interfaces[82]; and (3) they exhibit a low tendency to scavenge oxygen from BTO[83].

Epitaxially compatible high work-function electrodes for BTO often contain "stubborn" elements such as Ru, Pt, or Rh, which are difficult to evaporate at scale and tend to resist oxidation due to their high electronegativities (**top right, Figure 1e**), in contrast to more reactive elements like Sr (**bottom left, Figure 1e**). We focus on the growth of a preferred electrode candidate in **Figure 1b**, the perovskite oxide SRO, in an all-epitaxial metal-ferroelectric BTO-metal heterostructure. In conventional MBE, evaporation of Ru and oxidation to $Ru^{4+}$ are achieved by electron-beam evaporators and strong oxidants. These modifications enable adsorption-controlled growth[84]—a regime that promotes self-regulated stoichiometry and yields some of the lowest defect densities in SRO thin films[84–87]. Hybrid MBE[88] offers an alternative by replacing elemental Ru with the metal-organic precursor $Ru(acac)_3$, reducing the source temperature from ~ 2000 °C to < 200 °C[89] and preserving adsorption-controlled growth without the operational challenges of electron-beam evaporators and aggressive oxidants[90]. Similarly, BTO grown by hybrid MBE using titanium tetraisopropoxide (TTIP) as a Ti source exhibits adsorption-controlled growth windows[91–93] at practical growth rates and lower temperatures than predicted by conventional phase diagrams for titanates[94]. Overall, hybrid MBE is well-suited for achieving the desired stoichiometry[90–93], surface morphology[90,92], and optical properties[93] in BTO/SRO heterostructures, so we employ this technique in our work.

**2.2. $SrRuO_3$/$BaTiO_3$/$SrRuO_3$ heterostructures: Hybrid MBE growth and device fabrication**

We grew SRO/BTO/SRO heterostructures on 0.5 wt% Nb-doped SrTiO$_3$ (001) substrates (Nb:STO), with nominal thicknesses of 16 nm for each SRO layer and 40 nm for the BTO layer (**Figure 2a, left panel**), using hybrid MBE[89–92]. The SRO layers were grown at a substrate thermocouple temperature ($T_{sub}$) of ~ 700 °C, while the BTO layer was deposited at a higher $T_{sub}$ of ~ 950 °C. During SRO growth, strontium (Sr) was supplied at a beam equivalent pressure (BEP) of ~ 3 × 10$^{-8}$ Torr with an effusion cell, while ruthenium (Ru) was introduced by sublimating Ru(acac)$_3$ at a thermocouple temperature ($T_{Ru}$) of ~ 158 °C in a separate effusion cell. Growth conditions for SRO were optimized within an adsorption-controlled growth window to achieve the desired stoichiometry and surface morphology for heterostructure growth[90]. An inductively coupled radio-frequency oxygen plasma was maintained at 300 W and an oxygen partial pressure of ~ 8 × 10$^{-6}$ Torr throughout the growth process, including during temperature transitions between the growth of the BTO and SRO layers. For BTO growth, barium (Ba) was supplied at a BEP of ~ 3 × 10$^{-8}$ Torr using an effusion cell, while vapors of TTIP were introduced through a vapor inlet system at a BEP of ~ 4.36 × 10$^{-7}$ Torr, yielding a TTIP/Ba BEP ratio of 14.5 within a growth window for BTO[91–93].

Following growth, the heterostructures were patterned into circular capacitor devices with diameters ranging from 50 μm to 500 μm using reactive ion etching (RIE) with a BCl$_3$/Ar gas mixture (**Figure 2a, middle panel**). Platinum (Pt) contact pads were sputtered onto the top SRO electrode and the Nb:STO substrate to enable dielectric and ferroelectric measurements in a metal–insulator–metal geometry (**Figure 2a, right panel**). An optical micrograph of the fabricated capacitors (scale bar of 500 μm) is shown in the inset of **Figure 2b**. To minimize oxygen vacancy concentrations introduced during the etching process, the devices were annealed in 760 Torr of oxygen at 500 °C for 1 hour. Annealing at higher temperatures (700 °C) under similar oxygen conditions was found to cause irreversible Ru loss and the appearance of a SrO (002) peak (**Figure S1, Supplementary Information**), consistent with the observations of Shin *et al.*[95] (summarized in **Table S1, Supplementary Information**). Further experimental details on the hybrid MBE growth and device fabrication are provided in the **Experimental Methods**.

**2.3. Characterization of structural, dielectric, and ferroelectric properties**

High-resolution $2\theta$–$\omega$ coupled X-ray diffraction scans of the heterostructure, acquired both in the as-grown state and after device fabrication and post-growth annealing, confirm that the SRO and BTO layers are phase-pure and epitaxial (**Figure 2b**). Importantly, no signs of Ru loss or SrO formation were detected after annealing in an oxygen-rich environment at 500 °C, indicating thermal and chemical stability of the constituent layers under the chosen processing conditions. The sputtered Pt top contacts exhibit a (111) texture and are not epitaxially aligned with the underlying SRO electrode. To assess the strain state of individual layers and the tetragonality of the BTO film, we performed reciprocal space mapping (RSM) around the asymmetric (103) reflection of the Nb:STO (001) substrate (**Figure 2c**). The bottom SRO layer remains coherent with the substrate, as evidenced by its similar in-plane lattice parameter as the Nb:STO (identical *H*-position along the vertical dashed line in the RSM). This behavior is consistent with our expectations for a ~ 16 nm-thick SRO film grown on an STO (001) substrate[90]. In contrast, the BTO layer exhibits partial strain relaxation, with an in-plane lattice parameter $a_{\parallel,\text{BTO}}$ ~ 3.956 ± 0.005 Å and an out-of-plane lattice parameter $a_{\perp,\text{BTO}}$ ~ 4.044 ± 0.010 Å (marked with a + in **Figure 2c**). This strain state is intermediate between fully strained and relaxed *c*-domain BTO (summarized in **Table 1**) and corresponds to a tetragonality $a_{\perp,\text{BTO}}/a_{\parallel,\text{BTO}}$ of ~ 1.022 ± 0.004. This net elongation of the BTO unit cell along the out-of-plane direction (*c*-axis) favors the orientation of the polarization vector along the *c*-axis. Due to instrumental resolution limits, we do not resolve a separate (103) peak for the top SRO electrode. We expect the strain states of the top and bottom SRO electrodes to be different since they grow epitaxially on the partially-relaxed BTO layer and the Nb:STO substrate respectively. We estimate that the $(103)_{pc}$ peak of the top SRO layer if coherently strained to the BTO layer corresponds to an out-of-plane lattice parameter $a_{\perp,\text{SRO,top}} \sim 3.898$ Å (indicated by the star in **Figure 2c**), using a typical Poisson ratio $v = 0.3$ for a perovskite oxide[95]. This estimated position of the $(103)_{pc}$ peak of the top SRO electrode cannot be resolved from the (103) substrate peak due to the proximity of two peaks and the orders-of-magnitude higher intensity of the substrate peak.

We now address whether BTO retains ferroelectricity after device fabrication and annealing using a direct measurement of the polarization–voltage (*P*–*V*) response at different frequencies. We use a technique called Positive-Up-Negative-Down (PUND[96], described in the **Experimental Methods**) to separate the non-ferroelectric contributions to the current density from the ferroelectric contributions without assuming the functional forms or the frequency dependence of these deleterious non-ferroelectric contributions (unlike other compensation strategies discussed in **Section II (A), Supplementary Information**). A PUND voltage pulse train (schematic in **Figure 3a**) was applied using a fixed delay time $t_d = 10$ μs, a wait time $t_w = 10$ μs, and a voltage bias $V_{\text{bias}} = 0.5$ V. These parameters ensure that: (1) only the non-ferroelectric contributions decay between the P and U pulses (and N and D pulses), and (2) there is no observable back-switching due to the imprint effect (**Figures S2-S3, Supplementary Information**). We set equal rise and fall times $t_r = t_f$ in the pulse train (**inset of Figure 3a**), and measure the frequency-dependent current-voltage (dynamic *I-V*) response with PUND after establishing an equivalence with conventional four-quadrant dynamic *I-V* measurements at different frequencies (discussed in the **Experimental Methods**). All devices were independently measured in an interplanar capacitor configuration (**Figure 3b**), with the voltage bias applied to the top electrode and the bottom electrode grounded (a total of 4 devices where two of them have diameters of 150 µm and the other two have diameters of 200 µm). The normalized ferroelectric current densities for one of the 200 µm diameter devices (**Figure 3c**), obtained by subtracting the U and D current responses (non-ferroelectric) from the P and N current responses (total), exhibit clear switching peaks attributed to the ferroelectric behavior. The midpoint shifted to positive voltages, consistent with an imprint field arising from the structurally dissimilar electrodes suggested by RSM in **Figure 2c**. Integration of the ferroelectric current density over time yields hysteretic *P–V* curves (**Figure 3d**), providing direct evidence for ferroelectric switching with a non-zero remnant polarization in all-epitaxial SRO/BTO/SRO capacitors grown by hybrid MBE.

We quantify the ferroelectric response by extracting the remnant polarization $P_r$, coercive field $E_c$, and imprint voltages $V_{imprint}$ from PUND measurements across multiple frequencies (**Figure 3e**). Reported values represent averages over four devices, with standard deviations as error bars (**Tables S2–S5**, **Supplementary Information**). The remnant polarization remains similar at $P_r \sim 15.2$ µC cm$^{-2}$ over the measured frequency range of 500 Hz to 20 kHz. This value is ~ 5× the saturation polarization of the only previous report showing hysteretic *P-E* curves for MBE-grown BTO films[37] (where $P_r$ is not reported). Both $E_c$ and $V_{imprint}$ increase with frequency, consistent with frequency-dependent switching kinetics where higher fields are required to overcome slower dipole reorientation compared to the frequency of the alternating electric field[97]. Notably, the increase in $E_c$ and $V_{imprint}$ has a greater contribution from the switching peak in the negative-to-positive voltage sweep (**Figure 3c**). Taken together with the imprint effect observed in **Figures 3c** and **3d**, this observation suggests that switching from the up to the down polarization state is more difficult and *becomes increasingly more challenging* with increasing frequency than the polarization switching in the opposite direction. The observed asymmetry in switching dynamics may originate from (1) different strain states and crystalline qualities of the top and bottom electrodes (despite their nominally similar stoichiometry), (2) misfit dislocations formed during the partial strain relaxation of BTO[98] or (3) the presence of a gradient in defect dipole densities due to their alignment in the out-of-plane direction from compressive strain[99] and applied electric field[100].

To further investigate the bottlenecks in device performance we observed in PUND, we performed room-temperature impedance spectroscopy and static *I–V* measurements. Impedance spectroscopy (**Figure 3f**) shows a near-ideal capacitive response (with a phase angle $\theta \sim$ -90°) up to ~ 40 kHz—consistent with the frequency range where ferroelectric switching is observed in **Figure 3d**. A dielectric loss peak is observed at ~ $10^5$ Hz in the frequency range expected for dipolar relaxation processes[101,102]. The *I–V* characteristics (**Figure 3g**) exhibit a greater leakage current and resistance degradation under negative bias, suggesting that the top metal-ferroelectric interface limits the device performance. We now address the possible origin of the dipoles associated with the dielectric loss peak in **Figure 3f** and their distribution gradient, both of which may hasten resistance degradation near the top metal-ferroelectric interface in **Figure 3g**. One component of the defect dipole could be oxygen vacancies, which can arise from low oxygen partial pressures during MBE growth, plasma-induced damage during reactive ion etching, or insufficient annealing time[102]. These vacancies may act as negative charge traps which impede polarization switching and may redistribute along the electric field due to their higher mobility than cationic vacancies[103]. The other component of the defect dipole could be barium vacancies, which may form preferentially due to Ba re-evaporation[39] or ion milling damage[40]. To conclusively identify the source of resistance degradation and switching asymmetry, temperature-dependent *I–V* or impedance spectroscopy measurements could be employed to extract activation energies[104]. Complementary techniques sensitive to oxygen vacancies, such as electron spin resonance (ESR)[105], could further clarify the role of point defects in the dielectric and ferroelectric behavior of these heterostructures.

## 2.4. Benchmarking the performance of capacitor devices

We benchmark the ferroelectric properties of SRO/BTO/SRO heterostructures[18,25,38,40,44,45,48,52,60,62,63,68] by the choice of synthesis technique (**Figure 4a**) and substrate (**Figure 4b**). The $P_r \sim 15$ μC cm$^{-2}$ in our heterostructures is lower than those of similar heterostructures grown by PLD or sputtering, but within the same order of magnitude (**Figure 4a**) and with a similarly high $E_c$ as the heterostructures grown on STO (001) substrates (square symbols in **Figure 4b**). We note that the $P_r$ and $E_c$ of the heterostructures reported in **Figure 4** are also affected by the measurement conditions such as the area of the capacitor devices, where smaller capacitor areas may lead to better device performance. Our circular capacitor devices in this study have diameters ranging from 150 μm to 200 μm, while the majority of the devices shown in **Figure 4** have areas 1-3 orders of magnitude less than our devices, which may contribute to the observed differences in $P_r$ and $E_c$. We conclude thaat growth on the more closely lattice-matched scandate substrates (triangles in **Figure 4b**) and improved device fabrication protocols to counter the bottlenecks in switching dynamics and resistance degradation may be critical to enhancing device performance.

## 3. Conclusion

We have demonstrated the integration of ferroelectric BTO with epitaxial SRO electrodes using hybrid MBE, addressing key challenges in synthesizing lattice-matched metal–ferroelectric–metal heterostructures with Schottky barriers at both metal-ferroelectric interfaces. Hybrid MBE enables adsorption-controlled growth of both SRO and BTO, offering a scalable alternative to conventional MBE for these complex oxides and their heterostructures. We report device fabrication and annealing protocols that preserve the integrity of the heterostructure. We provide definitive evidence of ferroelectricity with hysteretic *P-E* curves using PUND, resulting in a $P_r$ that is 5× the saturation polarization of the only other report of hysteretic *P-E* curves in MBE-grown films[37]. The presence of an imprint and the asymmetric switching dynamics suggests differences in the metal-ferroelectric interfaces due to partial strain relaxation of BTO. We also hypothesize the presence of defect dipole gradients along the out-of-plane direction, likely driven by oxygen or barium vacancies and their migration in an applied electric field. These results establish hybrid MBE as a viable route for high-quality ferroelectric devices and offer key insights into defect-limited performance to guide the future design of devices with better performance.

## 4. Experimental Methods

### 4.1. Growth of SrRuO₃ and BaTiO₃ heterostructures with hybrid molecular beam epitaxy

The SrRuO₃/BaTiO₃/SrRuO₃ heterostructure was grown on a 0.5 wt% Nb:SrTiO₃ (001) substrate (Nb:STO, Crystec GmbH, Germany) in an EVO 50 MBE system (Scienta Omicron, Germany) using hybrid molecular beam epitaxy[88] for the SRO[89,90] and BTO layers[91–93]. The substrate was heated to a temperature ($T_{sub}$) ~ 700 °C and the *A*-site (*A* = Sr, Ba) fluxes were recorded as beam equivalent pressures (BEPs) using a beam flux monitor. The substrate was then cleaned for 20 minutes before the start of film growth with an inductively coupled radio-frequency oxygen plasma (Mantis, UK) operated at 300 W and at an oxygen pressure of ~ $8 \times 10^{-6}$ Torr. These oxygen plasma conditions were used throughout the heterostructure growth.

Both the top and bottom electrode layers of SrRuO$_3$ were grown at $T_{sub}$ ~ 700 °C. Strontium (Sr, 99.99%, Sigma-Aldrich, USA) was supplied at a beam equivalent pressure (BEP) of 3 × 10$^{-8}$ Torr using an effusion cell (MBE Komponenten, Germany). A solid metal-organic precursor, Ru(acac)$_3$, (99.99%, American Elements, USA) was used in place of elemental ruthenium (Ru) and sublimed from an effusion cell at thermocouple temperature ($T_{Ru}$) ~ 158 °C after identifying stoichiometric conditions within an adsorption-controlled growth window and the Ru-poor/stoichiometric boundary of the growth window which shows a higher tendency for terraced surfaces[90]. The barium (Ba) effusion cell was idled at a temperature of 300 °C in preparation for the growth of BaTiO$_3$ and to minimize its contribution to the background pressure during SrRuO$_3$ growth.

After the growth of the bottom SrRuO$_3$ electrode, the substrate was heated to $T_{sub}$ ~ 950 °C with the SrRuO$_3$ surface exposed to oxygen plasma. Barium (Ba, 99.99%, Sigma-Aldrich, USA) was supplied at a beam equivalent pressure (BEP) of 3 × 10$^{-8}$ Torr using an effusion cell (MBE Komponenten, Germany). Titanium tetraisopropoxide (TTIP, 99.999%, Sigma-Aldrich, USA) was used as a liquid metal-organic precursor for supplying titanium (Ti) using a vapor inlet system interfaced to the growth chamber through a gas injector (MBE Komponenten, Germany). The BEP of TTIP was ~ 4.36 × 10$^{-7}$ Torr, resulting in a TTIP/Ba BEP ratio of 14.5 which produced nominally stoichiometric BaTiO$_3$. The Sr and Ru effusion cells were idled at ~ 250 °C and ~ 60 °C respectively during the growth of BaTiO$_3$.

Following the growth of the BaTiO$_3$ layer, the heterostructure was cooled to $T_{sub}$ ~ 700 °C with the BaTiO$_3$ surface exposed to oxygen plasma in preparation for the growth of the top SrRuO$_3$ electrode. After the growth of the top SrRuO$_3$ electrode layer, the heterostructure was cooled with the SrRuO$_3$ surface exposed to an oxygen plasma operated at 250 W and at an oxygen pressure of

~ 5 × 10⁻⁶ Torr. The thickness of each layer is estimated from a calibration of the adsorption-controlled growth window for SRO and BTO.

### 4.2. Fabrication of defined capacitors

The SRO/BTO/SRO heterostructure was fabricated into defined circular capacitors with diameters ranging from 50 μm - 500 μm by photolithography with a positive photoresist (AZ 1518) and reactive ion etching. The etching step was performed in a reactive ion etcher (Oxford Instruments) with a mix of 20 sccm of $BCl_3$ and 5 sccm of Ar and an inductively coupled plasma operated at 1250 W power in 2 mTorr pressure. The heterostructure was etched to the Nb:STO (001) substrate, with an etch time of 5 minutes at an etch rate of ~ 20 nm/minute. Etching was followed by a second photolithography step (positive photoresist, AZ 1518) to pattern contact pads of smaller diameters on the top SRO layers. Platinum (Pt) was sputtered (AJA International) as top and bottom contacts with thicknesses of ~ 66 nm and ~ 33 nm respectively with an Ar plasma (20 sccm Ar at a pressure of 5 mTorr with 250 W power).

### 4.3. X-ray diffraction and reciprocal space mapping

The composition of the heterostructures in the as-grown state and after device fabrication and oxygen annealing was determined using high-resolution X-ray diffraction in a SmartLab XE diffractometer (Rigaku, USA) equipped with a two-bounce Ge (220) monochromator. The reciprocal space maps were acquired after device fabrication about the asymmetric (103) peak of Nb:STO between incident angles $\omega$ of 15° – 21° and reflected angles $2\theta$ of 70.5° – 78° to capture the coherently strained states and relaxed states of SRO and BTO. The scattering vectors along the in-plane ($H$) and out-of-plane ($L$) directions are expressed in relative lattice units in reciprocal space where the (103) peak of the substrate is at ($H, L$) = (1, 3).

### 4.4. Dynamic *I-V* measurements using the Positive-Up-Negative-Down (PUND) method

The PUND measurements were performed using the waveform generator module on a Keysight B1500A semiconductor analyzer. A PUND pulse train has five pulses (**Figure 3a**), a poling pulse towards negative voltages, two consecutive pulses towards positive voltages called the "Positive" (P) and "Up" (U) pulses, and two consecutive pulses towards negative voltages called the "Negative" (N) and "Down" (D) pulses. For a ferroelectric material, P and N (the "switching pulses") capture the total current response. If the delay time between consecutive pulses $t_d$ is chosen such that only the non-ferroelectric components of the current response decay, the U and D (the "non-switching pulses") exclusively capture the non-ferroelectric contribution to the current response.

As a first step, we design a PUND pulse train that is equivalent to a four-quadrant sweep ($V_{peak} \to 0 \to -V_{peak} \to 0 \to V_{peak}$) with a triangular waveform at a frequency $f$. We check that the circuits used are equivalent by checking that the direction of applied electric field is similar to the polarization direction in both measurements. Given that there are two "switching pulses" in a PUND pulse train, we expect the measurements to be equivalent when the frequency of the four-quadrant current-voltage measurements and the rise and fall times $t_r$ and $t_f$ of the PUND measurements are related by:

$$f = \frac{1}{2(t_r + t_f)}$$

The wait time $t_w$ at the peaks of the individual pulses changes the shape of the pulse from a triangle ($t_w \to 0$) to a rectangle ($t_w \gg t_r, t_f$). We use equal delay and wait times $t_d = t_w = 10$ μs and equal rise and fall times $t_r = t_f$ for all our measurements. For data acquisition, we use different step sizes and number of data points for the different components of the pulse train, collecting more points along the rise and fall times (50 data points each) than the delay (2 data points) and wait times (10 data points). Within each component of the pulse train, the data points are equally spaced. A voltage bias $V_{bias} = 0.5$ V was applied to prevent back-switching while accounting for imprint effects.

The ferroelectric contribution to the current density in the two sweeps extracted from P – U and N – D when integrated with respect to time give the polarization correct to a constant of integration. We shift the polarization values such that the magnitudes of the saturation polarization at both peak voltages are similar. We report the remnant polarization ($P_r$), coercive field ($E_c$), and imprint voltage ($V_{imprint}$, the horizontal shift in the center of the *P-E* hysteresis curve due to differences in the top and bottom electrodes) using the equations below:

$$P_r = \left. \frac{P_f|_+ - P_f|_-}{2} \right|_{V=0}$$

$$E_c = \left. \frac{V_c|_+ - V_c|_-}{2t_{BTO}} \right|_{P_f=0}$$

$$V_{imprint} = \left. \frac{V_c|_+ + V_c|_-}{2} \right|_{P_f=0}$$

### 4.5. Impedance spectroscopy measurements

Impedance spectroscopy measurements were performed on the defined capacitor devices using a Keysight E4990A impedance analyzer following the protocol outlined by Yang *et al*[106]. All calibration and measurement steps were performed in a two-terminal configuration. The impedance analyzer was calibrated with an open circuit, a short circuit, and a standard 100 Ω resistor (Keysight E4990-61001) as a device under test to account for contributions from the probe-station and the associated cabling after setting an oscillation level of 50 mV. We measure the standard 100 Ω resistor after the calibration to assess the frequency range where the device under test behaves as an ideal resistor (phase angle $\theta \sim 0°$) with a resistance of 100 Ω and conclude that the measurements are reliable between 20 Hz – $10^6$ Hz.

### 4.6. Static *I-V* measurements

*I-V* measurements were performed using the in-built *I-V* sweep module on a Keysight B1500A semiconductor analyzer using source measurement units (SMUs) with a noise floor of 100 fA. The voltage was swept on the top electrode from -2 V to 2 V to -2 V in a four-quadrant sweep, with a hold time of 3 s. The bottom electrode was grounded during this measurement.

**Acknowledgements**

The authors thank Dr. Stefano Gariglio for helpful discussions on polarization vs. electric field measurements. Film growth and characterization (A.K.M. and B.J.) was supported by the U.S. Department of Energy (Award No. DE-SC0020211). This work also benefitted from the Vannevar Bush Faculty Fellowship (Award No. N00014-19-1-2623) and the Air Force Office of Scientific Research (AFOSR, Award No. FA9550-23-1-0093). Z.Y. was supported partially by the UMN MRSEC program (Award No. DMR-2011401). Parts of this work were carried out in the College of Science and Engineering Characterization Facility at the University of Minnesota which receives partial support from the National Science Foundation through the Materials Research Science and Engineering Center (MRSEC, Award No. DMR-2011401) and the National Nanotechnology Coordinated Infrastructure (NNCI, Award No. ECCS-2025124) programs. A.K.M. also acknowledges the University of Minnesota Doctoral Dissertation Fellowship.

**Author contributions**

A.K.M., R.D.J., and B.J. conceived the idea and designed the experiments. A.K.M. grew the SRO/BTO/SRO thin film heterostructures and characterized the structural properties using X-ray diffraction and reciprocal space mapping. A.K.M. and Z.Y. fabricated the defined capacitors, assessed the suitability of different oxygen annealing conditions, and performed impedance spectroscopy. A.K.M., C.-H.L., and J.W. performed frequency-dependent four-quadrant current vs. voltage measurements under the direction of S.J.K. A.K.M. performed the PUND and static *I-*

*V* measurements and analyzed the data. A.K.M. and B.J. wrote the manuscript. All authors contributed to the discussion of the results and reviewed the manuscript.

**Conflict of interest**

The authors declare that they have no conflict of interest.

**Data availability**

All data needed to evaluate the conclusions of the paper are present in the paper and/or the Supplementary Information.

**Supplementary information**

The details of the choice of oxygen annealing conditions, dynamic *I-V* measurements using PUND, common sources of measurement artifacts, and detailed results of PUND measurements can be found in the Supplementary Information.

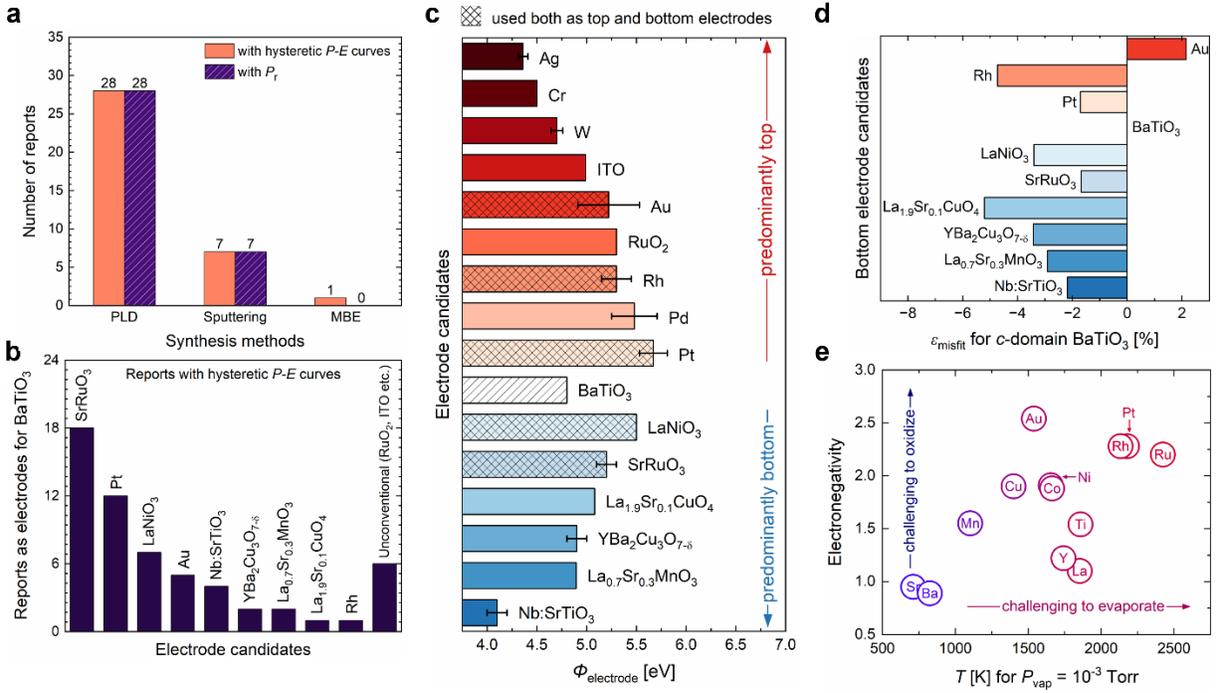

**Figure 1:** Electrode candidates demonstrated in metal-ferroelectric BaTiO$_3$-metal heterostructures[18,25,37–71]. **(a)** The number of reports (noted above each bar) of hysteretic *P-E* curves (solid bar, left) and those with a non-zero remnant polarization ($P_r$, solid bar shaded with diagonal lines, right) reported in metal-ferroelectric BaTiO$_3$-metal heterostructures by synthesis method. (b) Number of reports of different electrode candidates used with BaTiO$_3$ where hysteretic *P-E* curves are observed. Only unique electrode candidates are counted in each report and top and bottom electrodes are considered equivalent. **(c)** Work function of electrode candidates organized by their predominance as epitaxial bottom electrodes (blue, below the BaTiO$_3$ bar) and top electrodes (red, above the BaTiO$_3$ bar). The checkered pattern on the bar indicates that the candidate has been demonstrated as both a top and bottom electrode. Work functions of electrode candidates are reported from different sources outlined below. Work functions for single crystalline sources (bulk single crystals or epitaxial thin films) along the (001)$_{pc}$ facet are noted for Pt, Pd, Au, Rh, Ag, and W[72,73], SrRuO$_3$[74], Nb:SrTiO$_3$[74], and La$_{1.9}$Sr$_{0.1}$CuO$_4$[75]. Work functions for Cr[76], RuO$_2$[77], and LaNiO$_3$[78] are noted for polycrystalline samples. Work functions from crystalline films (where the facet is not reported) are noted for ITO (sputtered on *n*-type Si (001))[79], YBa$_2$Cu$_3$O$_{7-\delta}$ (deposited by pulsed laser deposition)[80], and La$_{0.7}$Sr$_{0.3}$MnO$_3$ (epitaxial on LaAlO$_3$ substrates)[81]. The work function of BaTiO$_3$ (bar shaded with diagonal lines) from films sputtered on metallic Ag electrodes on glass substrates is noted for comparison[107]. **(d)** In-plane misfit strain $\varepsilon_{misfit}$ between *c*-domain BaTiO$_3$ and the relaxed bottom electrode candidates for epitaxial growth showing most electrodes candidates used exert a compressive strain (negative misfit) favorable for *c*-domain growth. **(e)** Electronegativity[73] (Pauling scale) vs. temperature[108,109] required to achieve a vapor pressure of 10$^{-3}$ Torr for the elements constituting the bottom electrode candidates showing that some elements can be challenging to evaporate and/or oxidize for deposition by molecular beam epitaxy.

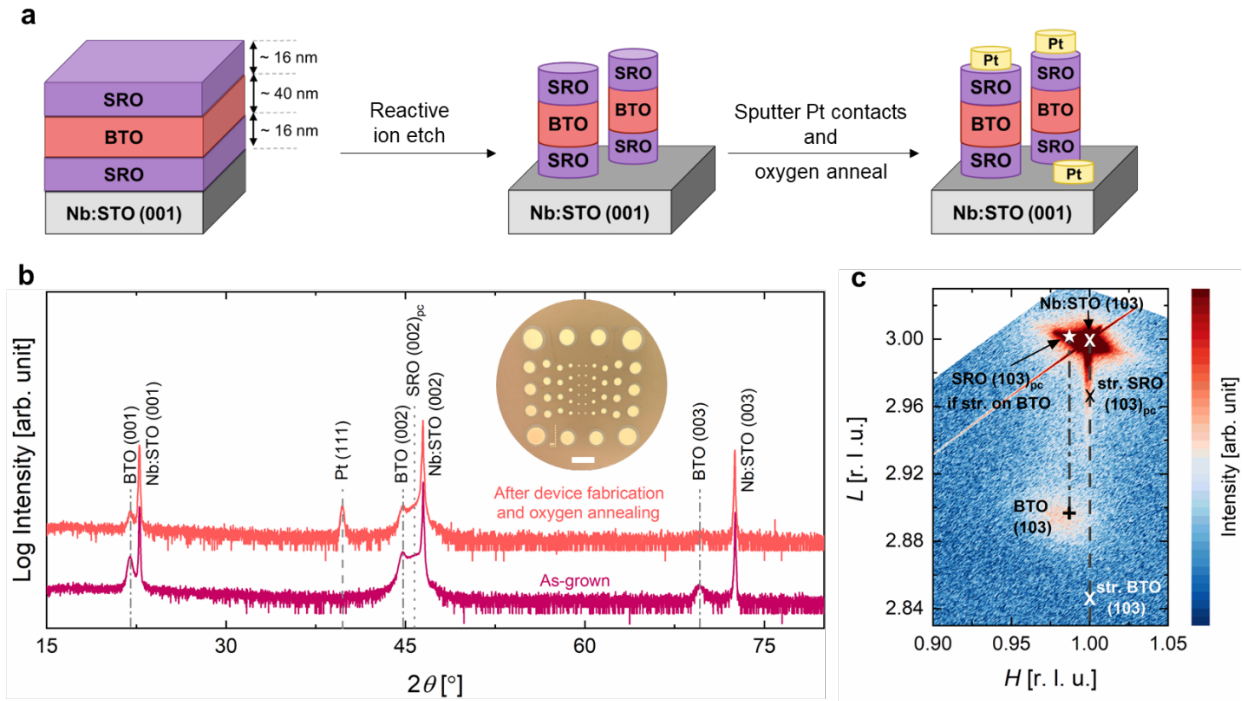

**Figure 2: (a)** Schematic showing the structure of the defined capacitor devices (thicknesses exaggerated for clarity) fabricated from the ~16 nm SRO/~ 40 nm BTO/ ~16 nm SRO/Nb:STO (001) heterostructure by reactive ion etching and Pt deposition for top and bottom contacts by sputtering. The devices were annealed in 760 Torr of oxygen at 500 °C for 1 hour. **(b)** $2\theta$-$\omega$ coupled X-ray diffraction scans for the as-grown heterostructure and after device fabrication and annealing (optical microscope image of the top view after device fabrication in the inset; scale bar is 500 µm) show phase pure and epitaxial SRO and BTO growth with no unintentional loss of Ru from annealing in an oxygen-rich environment to form SrO within the resolution of the diffractometer. Pt contacts are not epitaxial and are deposited in the (111) orientation. **(c)** Reciprocal space map about the (103) asymmetric peak of the substrate showing the bottom SRO is coherently strained to the Nb:STO (001) substrate (similar in-plane position along vertical dashed line at $H = 1$) and BTO is partially relaxed (+ symbol). A distinct peak for the top SRO layer could not be resolved. The expected peak position for this layer assuming coherent strain to BTO in the heterostructure is indicated by the star along the dash-dot vertical line from the BTO peak.

| Strain state | $a_{\perp,\text{BTO}}$ [Å] | $a_{\|,\text{BTO}}$ [Å] | $\dfrac{a_{\perp,\text{BTO}}}{a_{\|,\text{BTO}}}$ |
|---|---|---|---|
| Coherently strained BTO on Nb:STO (001) | 4.117 | 3.905 | 1.054 |
| Relaxed BTO | 4.036 | 3.992 | 1.011 |
| ~ 40 nm BTO film after device fabrication and oxygen annealing (this work) | 4.044 ± 0.010 | 3.956 ± 0.005 | 1.022 ± 0.004 |

**Table 1:** Lattice parameters in the out-of-plane ($a_{\perp,\text{BTO}}$) and in-plane direction ($a_{\|,\text{BTO}}$), and tetragonality ($a_{\perp,\text{BTO}}/a_{\|,\text{BTO}}$) of *c*-domain BTO in different strain states. The BTO layer in the SRO/BTO/SRO heterostructure after device fabrication and oxygen annealing shows an intermediate strain state between coherently strained *c*-domain BTO and relaxed *c*-domain BTO with a tetragonality greater than 1. The $a_{\perp,\text{BTO}}$ of coherently strained BTO is calculated using the in-plane misfit strain between BTO and Nb:STO and elastic coefficients of BTO from Bechmann *et al.*[110]

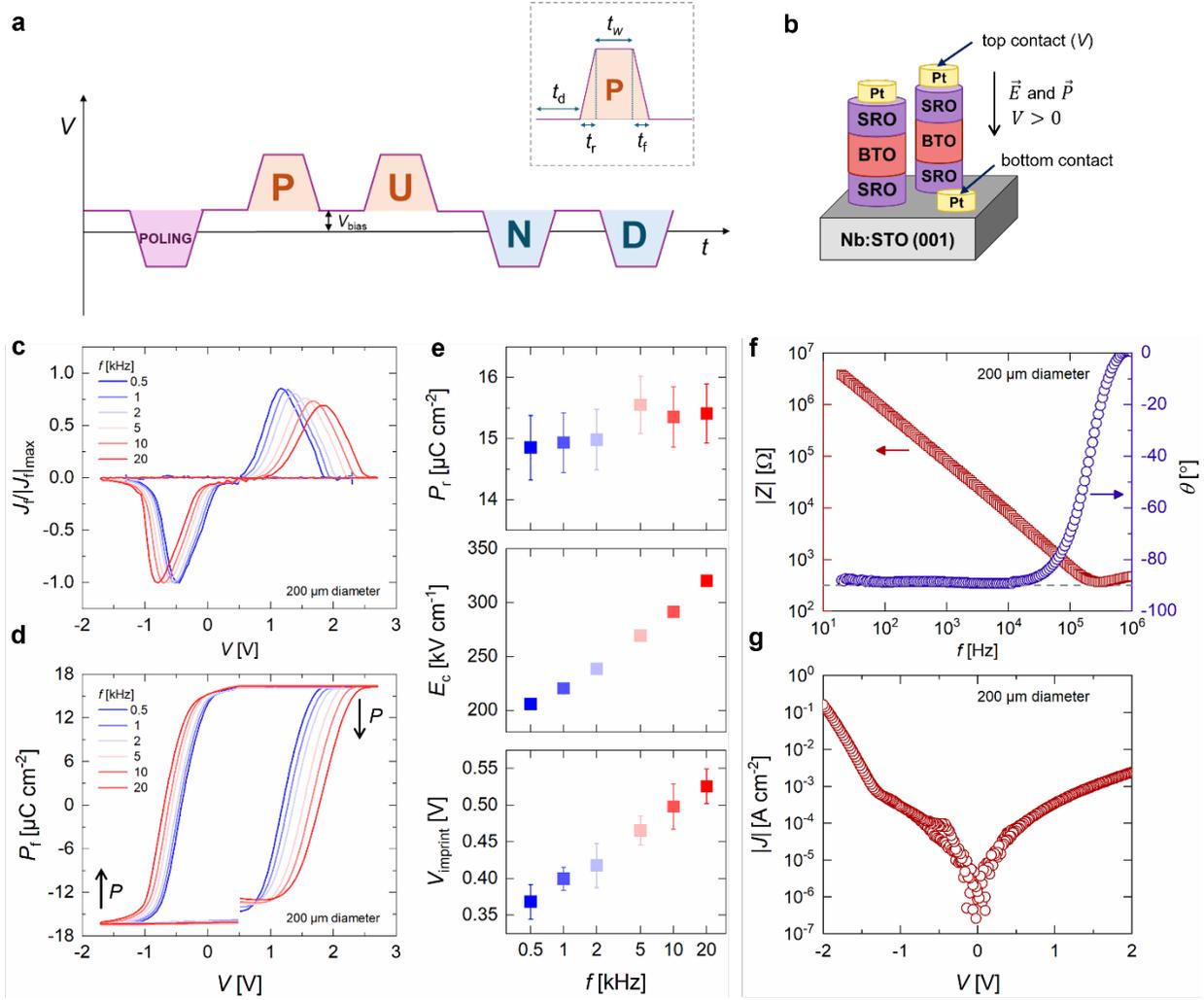

**Figure 3: (a)** Schematic of a PUND pulse train with five pulses: a negative poling pulse, "Positive" (P), "Up" (U), "Negative" (N), and "Down" (D) pulses in order. A pulse (inset) can be defined by four characteristic times – a delay time ($t_d$), a rise time ($t_r$), a fall time ($t_f$), and a wait time ($t_w$). A voltage bias $V_{bias}$ is used to shift the center of the pulse train to account for the horizontal shift in the center of the resultant polarization vs. voltage curves (the imprint effect) and to prevent back-switching. **(b)** Schematic of a device under test with the top Pt contact pad and bottom Pt pad on the Nb:STO (001) substrate after reactive ion etching used for the interplanar capacitor configuration. The direction of applied electric field and polarization is consistent for all measurements (PUND, impedance spectroscopy and static I-V) and points from the top electrode to the bottom electrode when $V > 0$ (the down polarization is accessed on saturation in the $V > 0$ direction). **(c)** Frequency-dependent normalized current density of the ferroelectric component vs. applied voltage for a device of 200 μm diameter after subtracting the non-ferroelectric components from the total current response using $t_d = t_w = 10$ μs and applying a voltage bias $V_{bias} = 0.5$ V to prevent back-switching. **(d)** Frequency-dependent polarization vs. applied voltage showing hysteresis expected for ferroelectric BTO for the same device. The polarization at saturation for the positive voltages is the down polarization (pointing from the top to the bottom electrode). **(e)** The remnant polarization ($P_r$), coercive field ($E_c$), and imprint voltage ($V_{imprint}$) as a function of frequency for four capacitor devices, two each of diameters 150 µm and 200 µm. The values are reported as the average of four devices and the error is the standard deviation. **(f)** Room-temperature impedance $Z$ and phase angle $\theta$ as a function of frequency for the 200 µm device in (c-d) showing ideal capacitor-like behavior ($\theta \sim$ -90°) up to $f \sim$ 40 kHz. **(g)** Four-quadrant static current vs. voltage curves for the 200 µm device in (c-d)

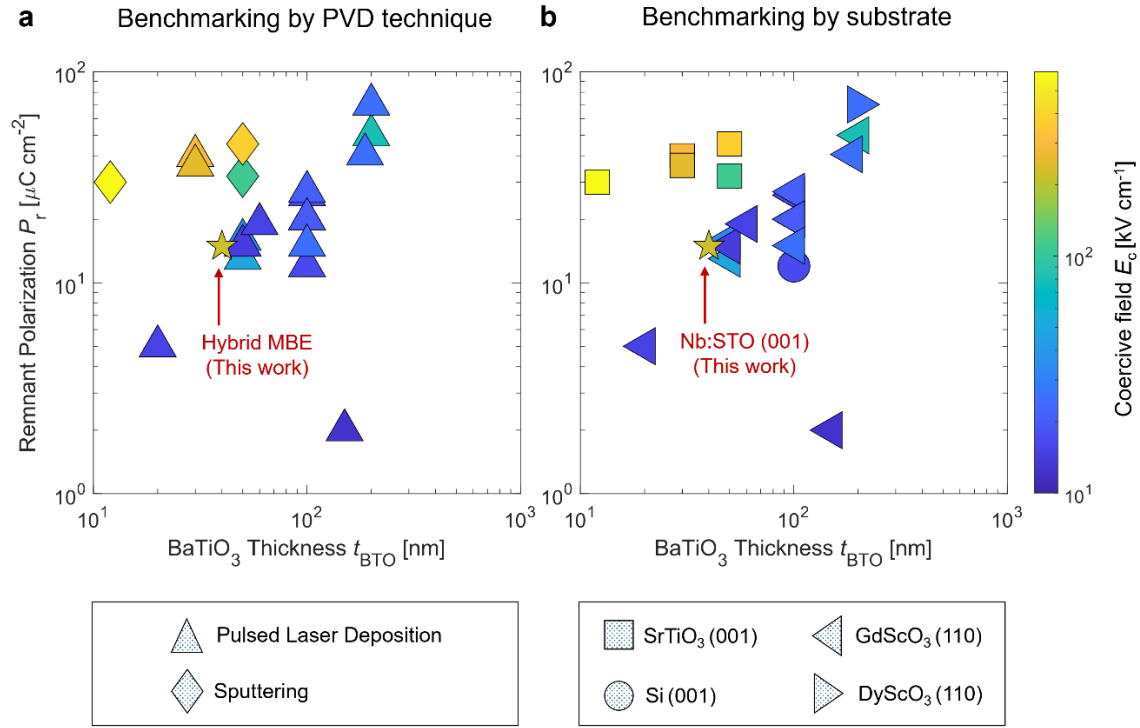

**Figure 4:** Benchmarking ferroelectric properties of SRO/BTO/SRO heterostructures by PVD technique and substrate[18,25,38,40,44,45,48,52,60,62,63,68]. Remnant polarization $P_r$ vs. BTO thickness $t_{BTO}$ is plotted for **(a)** different PVD techniques and **(b)** different substrates, with the symbol color as the coercive field $E_c$. The $P_r$ and $E_c$ at $f = 1$ kHz is reported for this work (star symbol). Heterostructures deposited by pulsed laser deposition are marked with upward-pointing triangles and those deposited by sputtering are marked with diamonds in (a). In (b), different symbols correspond to different substrates – square for SrTiO$_3$ (001), circle for Si (001), left-pointing triangle for GdScO$_3$ (110), and right-pointing triangle for DyScO$_3$ (110) substrates. The heterostructure grown on Si (001) substrates has an epitaxial buffer layer of SrTiO$_3$[25]. One report for SRO/BTO/SRO heterostructures grown by PLD has amorphous SRO as the top electrode and is recorded here for completeness[68].